\pdfoutput=1
\documentclass{webofc}
\usepackage[varg]{txfonts}   %
\usepackage{hyperref}
\usepackage{url}
\usepackage{enumitem}
\RequirePackage{orcidlink}   %
\hypersetup{colorlinks=true,citecolor=blue,urlcolor=blue,linkcolor=blue,
  pdftitle={AI in the EGI e-infrastructure -- Past experiences and future strategies},
  pdfsubject={CHEP 2026 proceedings},
  pdfkeywords={EGI Federation, artificial intelligence, AI in science, compute continuum, data continuum, federated LLM inference, interTwin, iMagine, AI Factories, RAISE}}
\graphicspath{{figures/}}
\begin{document}
\AddToHookNext{shipout/foreground}{\put(0.5\paperwidth,-1.4cm){\makebox[0pt]{\parbox[t]{\textwidth}{\centering\footnotesize\itshape Preprint of a paper submitted to the proceedings of the 28th Conference on Computing in High Energy and Nuclear Physics (CHEP 2026), to be published in EPJ Web of Conferences.}}}}

\title{AI in the EGI e-infrastructure -- Past experiences and future strategies}

\author{\firstname{Gergely}~\lastname{Sipos}\inst{1}\orcidlink{0000-0002-2148-3967}\fnsep\thanks{\email{gergely.sipos@egi.eu}} \and
        \firstname{Sergio}~\lastname{Andreozzi}\inst{1}\orcidlink{0000-0001-5567-4000} \and
        \firstname{Marco}~\lastname{Rorro}\inst{1}\orcidlink{0009-0002-6260-1007} \and
        \firstname{Andrea}~\lastname{Manzi}\inst{1}\orcidlink{0000-0001-7949-2199} \and
        \firstname{Tiziana}~\lastname{Ferrari}\inst{1}\orcidlink{0000-0003-3425-1409}
}

\institute{EGI Foundation, 1098 XG Amsterdam, The Netherlands}

\abstract{The EGI Federation has provided federated e-infrastructure services to European and global science for over 15 years. We review EGI's early experiences with AI in the interTwin and iMagine Horizon Europe projects, which built digital twins in physics and environmental sciences and an image-analysis platform for aquatic research. We then examine how the AI landscape has changed since 2023 and argue that the focus of AI in science has shifted from model training to adapting and applying AI to specific domains, projects and questions. Building on these lessons, we present the AI strategy emerging from the EGI Federation Strategy 2026--2030, organised around three enablers: (1)~an AI-ready infrastructure connecting compute, data and application resources across the continuum; (2)~human skills to know what is possible with AI and how to implement it; and (3)~trust, to run AI with confidence even in high-risk environments. We describe the projects implementing each enabler, including federated LLM inference, AI-ready data platforms and AI competence centres, and discuss a division of labour with EuroHPC's AI Factories, RAISE, research infrastructures and EOSC, in which EGI specialises in the customisation and long-term operation of AI for research.}

\maketitle

\section{Introduction}
\label{intro}

The EGI Federation, which emerged from WLCG in 2010, provides a federated e-infrastructure for over 160,000 researchers across all scientific disciplines. It is coordinated by the EGI Foundation and federates organisations through NGIs (national e-infrastructures) and scientific domain representatives, such as CERN for High Energy Physics or EMSO ERIC for marine and seafloor research.

The EGI infrastructure combines compute and storage resources with higher-level services, such as workload managers, container orchestration systems and Jupyter Notebooks, on top of which discipline-specific environments are operated. It is held together by `federation services', such as usage accounting, health monitoring, security oversight and user support, some of which EGI also delivers to WLCG, the European Open Science Cloud (EOSC) or Data Spaces. Within EC-funded projects, the EGI community also co-designs, validates and deploys new technologies into the infrastructure.

The ``EGI Federation Strategy 2026--2030''~\cite{strategy} is based on three pillars, with AI as a cross-cutting destination: (1)~Leadership in Sovereign Compute-Data Federation, (2)~A Model for Incentivised Cross-Border Access, and (3)~A Resilient and Sustainable Innovation Ecosystem.

Translated to AI, these pillars converge on three enablers that research performing organisations (RPOs) and research infrastructures (RIs) need for collective AI adoption: (1)~an AI-ready infrastructure, with connected compute, data and application resources; (2)~human skills, to know what is possible with AI and which implementation route to follow; and (3)~trust in AI, to run AI with confidence even in high-risk environments. Policy and funding instruments for cross-border access, the focus of the second pillar, underpin all three.

Section~\ref{sec-early} reviews EGI's early AI experiences, Section~\ref{sec-trends} summarises the trends that have since reshaped AI, Section~\ref{sec-strategy} details the three enablers and the projects implementing them, Section~\ref{sec-landscape} positions EGI among other European AI initiatives, and Section~\ref{sec-conclusions} concludes the paper.

\section{Early AI experiences before LLMs}
\label{sec-early}

\subsection{interTwin -- Digital twins in science}
\label{sec-intertwin}

The interTwin Horizon Europe project~\cite{intertwin} was coordinated by the EGI Foundation from 2022 to 2025. interTwin developed a prototype Digital Twin Engine (DTE) for science, and validated the engine with digital twin pilots in physics (radio astronomy, particle detector, gravitational wave) and environmental sciences (floods, cyclones, drought, wildfire, etc.).

The DTE had two key features: (1)~the ability to build and operate AI-powered digital twins that mimic the behaviour of complex systems, and (2)~the ability to use hybrid computing infrastructures composed of high-performance computing (HPC) and cloud systems. The \emph{itwinai} toolkit, a Python library, enabled both features within the DTE.

The modular architecture and standard interfaces of \emph{itwinai} allow users to scale workloads efficiently from laptops to supercomputers, reducing implementation overhead and improving resource utilisation. Several interTwin use cases developed AI models for their digital twins using hyperparameter optimisation performed on HPC systems of the project partners: the EuroHPC VEGA system in Slovenia and JUWELS at the J\"ulich Supercomputing Centre in Germany. The main project outputs, the DTE and its core \emph{itwinai} component, were later carried forward into other Horizon Europe projects, such as RI-SCALE, to increase their technology readiness level and broaden their use to more diverse AI scenarios.

\subsection{iMagine -- Image analysis in aquatic sciences}
\label{sec-imagine}

The iMagine Horizon Europe project~\cite{imagine} was coordinated by the EGI Foundation from 2022 to 2025. iMagine developed an operational AI platform and framework for image analysis in aquatic science, and validated the platform with 17 use cases across marine and freshwater ecosystems. All the use cases required image segmentation, object detection and classification in areas such as plankton and phytoplankton taxonomy, floating marine litter assessment, satellite-based oil spill tracking, coastal beach dynamics and freshwater diatom identification.

The iMagine platform had two key features: (1)~support for the entire AI workflow (data labelling, model training and fine-tuning, performance validation, model publishing and end-user inference); and (2)~the execution of these tasks on federated cloud computing sites of the EGI infrastructure (CSIC-IFCA in Spain, INCD in Portugal, TUBITAK-ULAKBIM in Turkey and Walton Cloud in Ireland).

The iMagine AI Platform was based on the \emph{AI4OS} software stack. Its modular architecture, standardised REST APIs, and integrated annotation and deployment tools allowed domain researchers to efficiently scale image processing workloads, reduce machine learning implementation overhead and streamline service deployment without needing deep AI engineering expertise. The project received Virtual Access funding to deliver the new AI-powered services to end-user scientists. Several resulting services have been running in production for years. For example, since August 2023 the OBSEA observatory of EMSO ERIC has used an iMagine model to detect and classify fish species in its video streams. The iMagine outputs were carried forward into the `European Digital Twin Ocean' EOSC Node and the Fluid-AI project (Section~\ref{sec-skills}).

\section{AI mega-trends since 2023}
\label{sec-trends}

When interTwin and iMagine were designed, applying AI in science largely meant training bespoke models on curated, labelled datasets. Since 2023, five mega-trends have rewritten this playbook:

\begin{enumerate}[leftmargin=*,itemsep=2pt]
\item \textbf{Reusable models}: high-quality foundation and domain models are now broadly available, many with open weights. Training a model is rarely where an AI project in science should begin.
\item \textbf{Customisation shift}: new customisation methods, such as parameter-efficient fine-tuning, retrieval-augmented generation (RAG), and prompt and skill engineering, make domain adaptation cheap. A few GPUs now do what once took a cluster.
\item \textbf{Grassroots initiatives}: researchers experience the value of AI in pilots and personal experiments. The challenge has shifted to moving from individual use to community-wide adoption.
\item \textbf{Agentic AI}: AI is moving from single-shot answers to autonomous, long-running tasks, in which trust and cost hinge on every step in a chain of actions.
\item \textbf{Rising memory and storage costs}: prices of memory and storage hardware are at record highs, so capacity needs careful planning and sharing.
\end{enumerate}

As a consequence, the focus of AI in science has shifted from model training to adapting and applying AI together, across institutions, disciplines and countries. For a federation such as EGI this is an opportunity: several lessons from interTwin and iMagine (the need for a compute continuum, integration into existing RI pipelines, API-first delivery and human support) become more relevant as the bottleneck moves from building models to customising, operating and trusting them at scale.

\section{EGI AI strategy: three enablers}
\label{sec-strategy}

\begin{figure}[t]
\centering
\includegraphics[width=\textwidth]{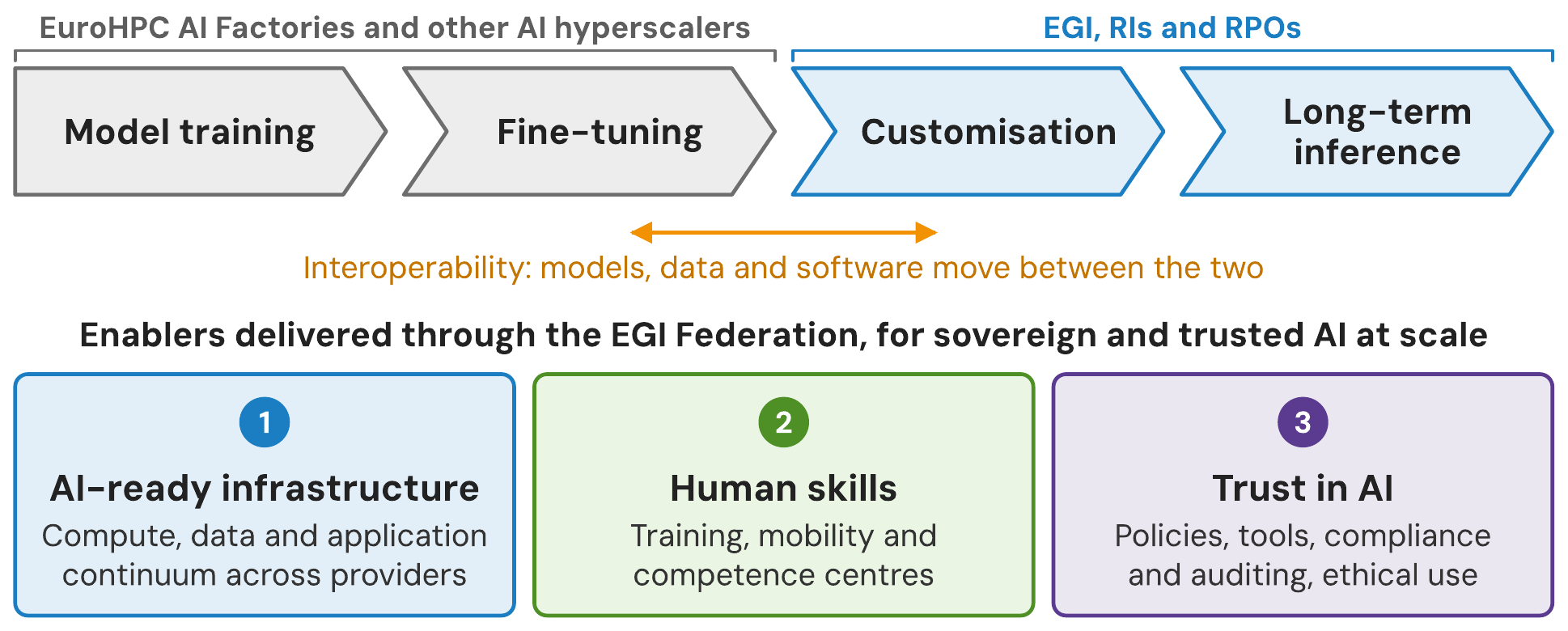}
\caption{The three enablers of the EGI AI strategy, and the division of labour along the AI lifecycle between the EuroHPC AI Factories on one side and EGI, RIs and RPOs on the other (see Section~\ref{sec-landscape}).}
\label{fig-enablers}
\end{figure}

The EGI Federation Strategy 2026--2030~\cite{strategy} defines priorities for the EGI community for the next five years, with AI as a cross-cutting destination. Building on the early experiences and the trends introduced above, the strategy translates to AI as three enablers that EGI brings together for research performing organisations and research infrastructures: an AI-ready infrastructure, human skills and trust in AI (Figure~\ref{fig-enablers}). The enablers are complementary: infrastructure without skills remains underused, while infrastructure and skills without trust cannot reach high-risk and high-value scientific applications. The next subsections describe each enabler and the projects that implement it.

\subsection{AI-ready infrastructure}
\label{sec-infra}

Achieving `Leadership in Sovereign Compute-Data Federation' for AI means connecting compute, data and application resources across organisations and communities. This has three practical consequences: extending the compute continuum, building a data continuum and establishing an application continuum of AI services. The following subsections describe how each is being implemented.

\subsubsection{Compute continuum}
\label{sec-compute}

The EGI infrastructure started as a network of high-throughput computing (HTC) clusters suited to large-scale parameter-study applications (many independent computational tasks). In 2012, the infrastructure broadened its scope and introduced `Cloud compute' resources, which widened the range of computational tasks that EGI can run to include interactive notebooks, hosting of long-running services, virtualised native applications and graphical visualisation. The compute service portfolio was further expanded in 2022 with the introduction of the `Container compute' service that simplified application porting across in-house and external platforms.

In the same year, the EGI community started working on a federated HPC service that would connect national HPC systems to the EGI Federation, further expanding the compute continuum and making it suitable for AI model training, multi-scale modelling workflows and other compute-intensive jobs. This EGI HTC-cloud-container-HPC compute continuum already federates GPU resources from seven sites, and the number keeps growing thanks to national investments going into compute infrastructures and European projects facilitating their integration into EGI. Beyond model training, the goal of the continuum is to enable cross-national model portability and inference, as well as `frugal AI', i.e.\ running AI workloads on the most appropriate and most energy-efficient resources. The GreenDIGIT and ENSURE projects contribute to the latter with methods and tools to measure and reduce the environmental footprint of computing, while the EGI HPC-AI Working Group coordinates the integration of national HPC and GPU capacity.

Two challenges remain in extending the compute continuum towards AI: (1)~a lack of well-defined use cases and sectors that genuinely require federating HPC and cloud resources across countries, such as federated learning, or training and inference on AI models using sensitive data that cannot leave its country of origin; and (2)~the absence of a sustainable business model for cross-national use, e.g.\ cost-recovery mechanisms, or workload managers that route jobs to the national resources where their users hold compute allocations.

\subsubsection{Data continuum}
\label{sec-data}

Data is even more essential in the AI era, given how much data model training and fine-tuning consume, and the scale at which scientific data can benefit from AI inference. EGI has so far offered rather basic data management solutions (e.g.\ data staging and storage mounting); it now needs to build a data continuum that connects scientific data repositories to AI applications.

This aligns with the desire of data repositories to increase the reuse of their data for AI, without becoming compute centres themselves. Two projects, both started in 2025, address this from different perspectives.

The RI-SCALE project~\cite{riscale} brings together repositories of European RIs (such as Euro-BioImaging, BBMRI, EISCAT, ENES) and EGI compute centres (TU Wien, TUBITAK) to establish `Data Exploitation Platforms' that stage data from long-term repositories to HPC/cloud systems, expose them to AI platforms and models, manage user access and remove unused data to reclaim resources. The platforms will be jointly operated by repository and compute providers, with costs and quotas based on measured resource consumption and environmental impact per user session.

EOSC Data Commons (EDC)~\cite{edc} helps national data repositories get their data used as broadly as possible, through two services:

\begin{enumerate}[leftmargin=*,itemsep=2pt]
\item A discovery system, Matchmaker, that harvests and enriches metadata, supports semantic search over vector embeddings, and uses large language models (LLMs) to rerank the results and summarise their relevance, covering both datasets and the applications that can analyse or visualise them.
\item An executor system, Data Player, that packages the chosen dataset and application pair, copies them to the user's compute environment and executes them.
\end{enumerate}

RI-SCALE (European RIs) and EDC (smaller, national repositories) build a foundational layer to connect data repositories and compute resources in EGI, complemented by domain-specific projects such as ENVRI-Hub NEXT~\cite{envrihub} for environmental science RIs, ECHOES~\cite{echoes} for cultural heritage, EUCAIM~\cite{eucaim} for cancer images and SAGE~\cite{sage} for a Green Deal Data Space.

\subsubsection{Application continuum}
\label{sec-apps}

A key challenge for researchers is keeping up with rapidly evolving AI technologies, models and software frameworks. To remain at the forefront for researchers, EGI must offer innovative, user-friendly AI environments that blend seamlessly into existing research practices (e.g.\ gateways that run inference using models hidden behind existing interfaces) or support the customisation of AI applications. EGI is working towards this by introducing new AI-oriented services in its portfolio: a Federated LLM service (FedLLM), acting as a `shared brain' across EGI sites, and an AI framework service (AI Lab) built from community contributions.

FedLLM focuses on federating the LLM inference services emerging from multiple EGI members. The independently operated inference services are harmonised through a common OpenAI-compatible API and web interface. EGI Check-in provides federated authentication and supports access policies and quotas, while LiteLLM acts as the common gateway and routing layer towards provider-side inference services. The initial FedLLM includes providers from the Czech Republic (Masaryk University) and Hungary (HUN-REN SZTAKI), offering six different models developed in Europe, the United States and China (including Kimi K3, an open-weight, 2.8-trillion-parameter model). The service is open to early adopters, both users and additional providers, with typical uses including metadata generation, coding assistance and research automation. One early adopter is Neurodesk.org, an open community of neuroscientists who integrated EGI FedLLM into their Jupyter Notebook environment as a built-in coding assistant and agent. The FedLLM service is also used as a core component within EOSC Data Commons, powering dataset and application discovery.

The AI Lab service targets the broader AI lifecycle and offers integrated environments to customise, deploy and operate models for inference. Its implementations are community contributions based on established and emerging solutions: the \emph{AI4OS} technology that was proven in EGI during the iMagine project (see Section~\ref{sec-imagine}), and the BioImage.IO framework and the \emph{itwinai} toolkit that are used in RI-SCALE. Additional implementations are expected to emerge from future domain-specific projects and partnerships.

Looking ahead, the growing adoption of agentic AI in science is expected to shape the service's future direction. Further AI Lab implementations may therefore extend beyond model customisation and inference to include platforms for agent development, customisation and operation, enabling researchers to build, adapt and run AI agents that support their scientific workflows, with the trust requirements discussed in Section~\ref{sec-trust}.

\subsection{Human skills}
\label{sec-skills}

The second enabler is human skills: knowing what is possible with AI and the best implementation route to follow, both within and outside EGI, complemented by support for infrastructure users. It builds on lessons already learned from projects such as iMagine. A new project, Fluid-AI (October 2026 to September 2029), establishes a core team of `advisors for adoption' that extends the iMagine Competence Centre model beyond its original scope, scaling it to multiple scientific domains (not only aquatic science), to multiple modalities (not only images), and to multiple problems (not only image segmentation and object detection), while also strengthening and simplifying the AI platform's direct connection to scientific databases. EGI aims to further expand this competence centre to additional technologies, AI use cases and disciplines.

Complementing this domain-specific competence-building, the recently started, EGI-led RItrain-DIGIT project (September 2026 to August 2029) addresses the technical skills gap at the infrastructure level: it offers generic and in-depth training on the use of national compute centres. One of its learning pathways is dedicated to using AI in science and will be delivered through online and face-to-face courses and with `mobility visits' that strengthen bilateral connections between domain research groups and their national compute centres.

\subsection{Trust in AI}
\label{sec-trust}

The third enabler, and EGI's next move, is trust: the confidence to run AI even in high-risk environments. The need for trust grows as AI in science progresses from augmented AI (assisting a researcher), through automated AI (executing well-defined steps of a workflow), to autonomous AI (agents planning and executing long-running tasks), and ultimately towards automated scientific discovery. Along this path three challenges become critical: compounding errors, as small inaccuracies propagate and amplify through chains of agent actions; the design and placement of human checkpoints, where researchers validate intermediate results and decisions; and reproducibility, since results produced by non-deterministic, frequently updated models and dynamic tool use are hard to repeat and verify.

Trust is equally a matter of policy, at several levels: institutional policies of research performing organisations, project- and experiment-level policies, and sectoral policies. Policies alone are insufficient, however: they require tools for implementation (e.g.\ provenance capture, model and dataset versioning, and guardrails in AI services), mechanisms for compliance and auditing, and practical guidance for ethical use. As a federation that already operates security, policy and accounting services across many providers, EGI is well placed to extend these federation services towards trusted AI, and plans to address these topics in future projects and partnerships.

\section{EGI in the European AI landscape}
\label{sec-landscape}

EGI develops its AI strategy alongside other European AI initiatives with complementary scopes, and a division of labour is emerging among them along the AI lifecycle (Figure~\ref{fig-enablers}): the EuroHPC AI Factories and other AI hyperscalers are best suited to large-scale model training and fine-tuning, while EGI, RIs and the research performing organisations they federate focus on customising models for specific scientific communities and on operating them for long-term inference, close to researchers and their data. Interoperability between the two worlds is essential for moving models, data and software between them.

Complementarity and co-existence with EuroHPC, and specifically with its AI Factories, is a critical topic for EGI and for the EGI members that have a foot in both infrastructures. RIs and EOSC are also developing their own AI activities, and these should be connected with the EuroHPC and EGI ecosystems too. Collaboration on this front is expected to intensify from 2027 thanks to new HORIZON-INFRA-TECH-01-01 and INFRA-2025-01-EOSC-05 projects.

At the European coordination level, SCIANCE (December 2025 to May 2028) is the foundational project of RAISE (the Resource for AI Science in Europe). It mobilises leading European scientific organisations and RIs to advise the European Commission on how to grow the RAISE initiative further. EGI is involved in the `infrastructure upgrade scenarios' activity of SCIANCE, and through it will advocate broadening the infrastructure pillar of RAISE to a balanced and inclusive approach that supports the entire value chain, from model training to long-term inference.

\section{Conclusions}
\label{sec-conclusions}

EGI's early experiences with AI, and the trends that have reshaped AI since 2023, point to a clear conclusion: Europe needs a holistic approach to AI in science. Infrastructure, skills and trust form a continuum, and investing in one without the others leaves value unrealised. Europe's competitive edge can come from specialisation-based interconnectivity: AI Factories focusing on model training and fine-tuning, EGI and its members on customisation and long-term inference. Interoperability for shipping models, data and software between these systems is a shared responsibility.

EGI is well suited to lead on the research-sector end of this chain. It has already made bold moves on the infrastructure layer, spanning compute, data and software (Section~\ref{sec-infra}); it has launched new initiatives to develop human skills through training and user support (Section~\ref{sec-skills}); and it is preparing future projects on responsible, trusted and ethical AI (Section~\ref{sec-trust}). These enablers map naturally onto the pillars of the European Commission's `AI Continent Action Plan' (April 2025)~\cite{aicontinent}, confirming that EGI's strategy contributes directly to the shared European vision for AI leadership. The main challenge over the five-year implementation of the strategy will be to fund these activities sustainably: first by securing the projects that start them, then by embedding them in the core institutional operations of EGI members, partners and the EGI Foundation, so that they no longer depend on individual project funding.

We have discussed how AI can be used in EGI to empower its users to achieve `better and faster' science. However, AI can benefit the EGI infrastructure operators too, to improve infrastructure performance and the speed of innovation, to prevent cybersecurity attacks, or to lower the overall effort for infrastructure maintenance and user support (as the AI assistant recently introduced in the EGI/WLCG Helpdesk shows~\cite{ggus}). The topic of `using AI for EGI operations' is outside the scope of this paper and will be revisited in a future publication.

\begin{acknowledgement}
This work was supported by the European Union's Horizon Europe research and innovation programme under grant agreements No.~101058386 (interTwin), No.~101058625 (iMagine), No.~101188168 (RI-SCALE) and No.~101292852 (RItrain-DIGIT).
\end{acknowledgement}

\bibliography{refs}

\end{document}